# Transport in the Laughlin quasiparticle interferometer: Evidence for topological protection in an anyonic qubit

F. E. Camino, W. Zhou and V. J. Goldman
*Department of Physics, Stony Brook University, Stony Brook, New York 11794-3800, USA*

We report experiments on temperature and Hall voltage bias dependence of the superperiodic conductance oscillations in the novel Laughlin quasiparticle interferometer, where quasiparticles of the 1/3 fractional quantum Hall fluid execute a closed path around an island of the 2/5 fluid. The amplitude of the oscillations fits well the quantum-coherent thermal dephasing dependence predicted for a two point-contact chiral edge channel interferometer in the full experimental temperature range $10.2 \leq T \leq 141$ mK. The temperature dependence observed in the interferometer is clearly distinct from the behavior in single-particle resonant tunneling and Coulomb blockade devices. The $5h/e$ flux superperiod, originating in the anyonic statistical interaction of Laughlin quasiparticles, persists to a relatively high $T \sim 140$ mK. This temperature is only an order of magnitude less than the 2/5 quantum Hall gap. Such protection of quantum logic by the topological order of fractional quantum Hall fluids is expected to facilitate fault-tolerant quantum computation with anyons.

## I. INTRODUCTION

Topological quantum computation with anyons has been suggested as a way of implementing intrinsically fault-tolerant quantum computation.[1-3] Intertwining of anyons, two-dimensional (2D) particles with non-integer exchange statistics,[4,5] adds a non-trivial Berry phase[6] to the system wave function, that is, induces unitary transformations of the many-particle wave function that depend only on the braiding order and on the mutual braiding statistics of the particles. The unitary transformations resulting from braiding of anyons can be used to perform quantum logic. The topological nature of such logic makes it robust against many sources of quantum decoherence, provided an appropriate restricted Hilbert space is chosen as the computational basis and the dynamical phase evolution is minimized during computation.

The only physical systems experimentally demonstrated to realize fractional-statistics particles are the gapped fractional quantum Hall (FQH) fluids.[7-9] The elementary charged excitations of a FQH fluid, Laughlin quasiparticles (LQPs), have been experimentally confirmed to have fractional electric charge[10] and anyonic braiding statistics.[11,12] Their anyonic statistics is determined by the topological order of the underlying FQH fluid,[13-15,9] and can be either commutative (abelian) or non-commutative (non-abelian), for various FQH fluids. Specific physical implementations for abelian and non-abelian qubits and logic gates based on quantum antidot and interferometer devices have been proposed.[3,16,17]

In this paper we report experiments on thermal broadening of the superperiodic oscillations in the novel LQP interferometer, illustrated in Fig. 1(a), where quasiparticles of the 1/3 FQH fluid execute a closed path around an island of the 2/5 fluid. The Aharonov-Bohm flux superperiod $\Delta_\Phi = 5h/e$ has been attributed to the mutual anyonic statistics of the FQH quasiparticles.[11,12,18] We observe the $5h/e$ superperiodic oscillations to persist to $T \sim 140$ mK, a temperature comparable to the 2/5 FQH gap, and much higher than the estimated LQP quantization energy in the edge confining potential.

A 2D electron system in a quantizing magnetic field forms effectively 1D chiral edge conducting channels, the direction of circulation being determined by the field.[19-21] The detailed dynamics of FQH edge channels is described by the chiral Luttinger liquid ($\chi$LL) models,[22-31] and is characterized by the Luttinger exponent $g$. We analyze the experimental data, the oscillatory dependence of differential conductance on dc Hall voltage, and the $T$-dependence of the oscillatory conductance amplitude, using the edge excitation velocity $u$ and $g$ as two fitting parameters. We find good agreement with the part of the $\chi$LL theory[26] describing the interferometric oscillations, provided $g$ is renormalized to ~0.8. The experimental bias and temperature dependencies of the average (non-oscillatory) conductance do not agree well with the predictions for $g = 1/3$ single-constriction LQP tunneling.

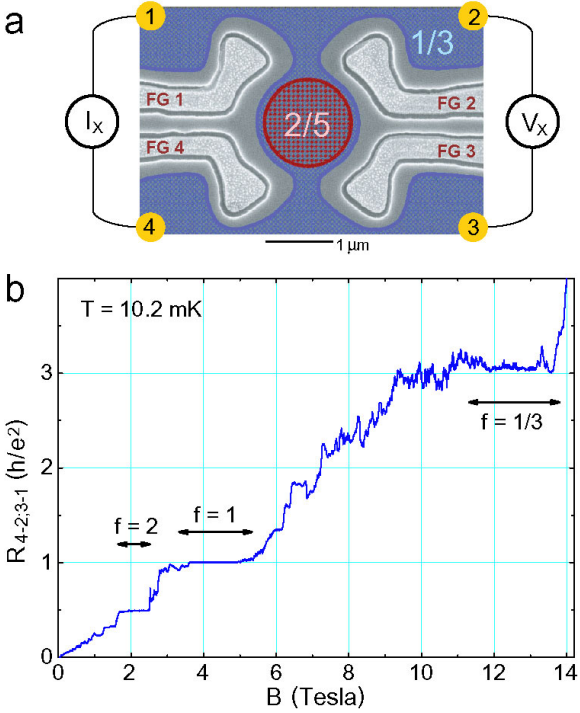

FIG. 1. (a) Scanning electron micrograph of a typical electron interferometer device. The front gates in shallow etch trenches define the central island of 2D electrons ~300 nm below the surface. The main depletion potential is produced by the etch trenches, the gates are used for fine-tuning the two wide constrictions for symmetry. The four numbered Ohmic contacts are in fact at the corners of a 4×4 mm sample. The filling 2/5 island is surrounded by the 1/3 FQH fluid, which forms the Aharonov-Bohm ring. (b) Four-terminal Hall resistance $R_{XY} \equiv V_{4-2}/I_{3-1}$ is determined by the quantum Hall filling $f$ in the constrictions, giving definitive values of $f$. The fine structure is due to quantum interference effects, including the conductance oscillations as a function of magnetic flux through the island.

## II. EXPERIMENTAL TECHNIQUE

The LQP interferometer sample, illustrated in Fig. 1(a), was fabricated as described before.[11,18,32] The chiral edge channels form at the periphery of the undepleted 2D electrons, following the constant electron density equipotentials around the etch trenches. Here we focus on the situation when a quantum Hall filling $f = 1/3$ annulus surrounds an island of the $f = 2/5$ FQH fluid, as shown schematically in Fig. 1(a). We are confident that the current is transported by the $f = 1/3$ fluid because the Hall resistance is quantized to $R_{XY} = 3h/e^2$, see Fig. 1(b). In this regime, we observe the superperiodic conductance oscillations as a function of magnetic field $B$, with the flux period $\Delta_\Phi = 5h/e$, Fig. 2(a), as has been reported.[11,12,18] The charge $e/3$ LQP consecutive orbits around the island are quantized so that the total Berry phase acquired in circling the island (sum of the Aharonov-Bohm and the anyonic contributions) is an integer multiple of $2\pi$.[33] The island center electron density is only ~4% less than the well-known 2D bulk density, and thus has essentially the same quantum Hall filling. The island filling is further confirmed by the ratio of the flux to the backgate oscillation periods, which is proportional to $1/f$, and was calibrated in the integer quantum Hall regime.[11,12]

The experiments were performed in a top-loading dilution refrigerator with an extensively RC-filtered sample and thermometer wiring. The Ruthenium oxide chip resistor thermometer has been calibrated by [60]Co nuclear orientation absolute thermometry for $T \leq 40$ mK, and against a commercially calibrated resistor for $T \geq 25$ mK. Relatively small thermometer magnetoresistance corrections at 12 Tesla (under 2 mK in the experimental range) were taken into account. The tunneling conductance has been measured using an $I_X = 93$ pA rms ac excitation at 5.4 Hz by a lock-in-phase detection technique. The resulting Hall voltage bias $V_H = R_{XY} I_X$ on the $R_{XY} = 3h/e^2$ plateau is 7.2 μV; adding incoherent "electromagnetic noise environment" incident on the sample[30] yields a total of $\sqrt{(7.2)^2 + (2)^2} = 7.5$ μV. The peak *tunneling* current (heating the interferometer active region) is ~0.8 pA, yielding the total interferometer heating power ~6×10$^{-18}$ W. The rest of the excitation and the "noise environment" power incident on the sample (~7×10$^{-16}$ W) heats the relatively large sample's Ohmic contacts, immersed in the $^3$He-$^4$He mixture bath.

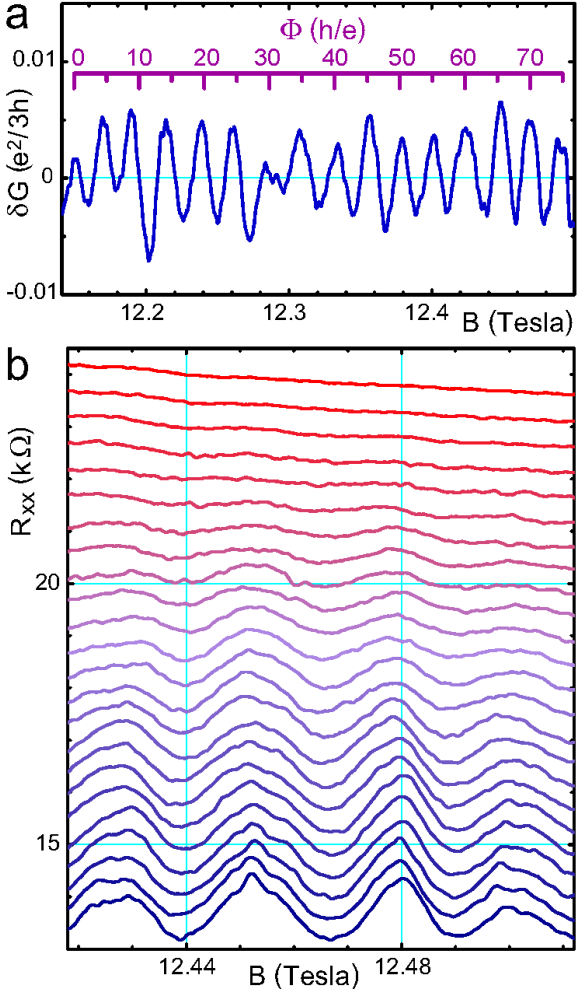

FIG. 2. Superperiodic conductance oscillations of Laughlin quasiparticles. (a) The magnetic flux period $\Delta_\Phi = 5h/e$ corresponds to creation of ten $e/5$ quasiparticles in the island.[11,33] (b) Evolution of the superperiodic oscillations in the $10.2 \leq T \leq 141$ mK temperature range. The successive traces are offset vertically by 0.4 kΩ. The oscillatory conductance $\delta G$ is calculated from the directly measured $R_{XX}$ as described in the text. The data in (a) and (b) were obtained on different cooldowns of the sample.

The temperature dependence of the oscillations in the directly measured $R_{XX} = V_X / I_X$ is shown in Fig. 2(b) for $10.2 \leq T \leq 141$ mK. The ~12.9 kΩ constant background results from the quantized longitudinal $R_{XX} = R_{XY}(\frac{1}{3}) - R_{XY}(\frac{2}{5}) = \frac{1}{2} h/e^2$, where $R_{XY}(\frac{1}{3})$ is the Hall resistance of the interferometer region, and $R_{XY}(\frac{2}{5})$ comes from the 2D bulk outside, connected to the Ohmic contacts.[20,23] These particular $R_{XX}(T)$ data were obtained continually over a ~70 hour period (following a ~100 hour sample stabilization period), which demonstrates the stability of the superperiodic conductance oscillations in this FQH regime.

### III. PHENOMENOLOGICAL ANALYSIS

In order to compare the experiment to a theory quantitatively, we analyze the raw data as follows. We determine the amplitude of each of the several regular oscillations, $\delta R_{XX}(T) = \frac{1}{2}(R_{XX,MAX} - R_{XX,MIN})$, at each temperature. The oscillatory conductance $\delta G$ is calculated from the directly measured $\delta R_{XX}(T)$ and the quantized value of the Hall resistance $R_{XY} = 3h/e^2$ as $\delta G = \delta R_{XX}/(R_{XY}^2 - \delta R_{XX} R_{XY})$, a good approximation for $\delta R_{XX} \ll R_{XY}$.[19-21,23,30] We then normalize $\delta G(T)$ to the average of the two lowest temperatures (10.2 and 11.9 mK) for each particular oscillation, and take the average of $\delta G(T)$ for six thus normalized oscillations, in order to reduce the experimental uncertainty. The normalized conductance amplitude $G_A(T)/G_A(11\,\text{mK})$ data are shown in Fig. 3 on the linear and semi-log scales. The experimental $G_A(T)$ varies by 31, when $T$ varies by 14. It is apparent in the semi-log plot that the higher-$T$ oscillation amplitude follows $G_A(T) \propto \exp(-T/T_0)$, the dependence expected for thermal dephasing of a quantum-coherent interference signal, with a surprisingly small $T_0 \approx 27$ mK, yet the oscillations persist up to 140 mK.

We first check whether the finite $V_H = 7.2$ μV ac excitation used to measure $G_A(T)$ in the lock-in technique is in the linear regime. Numerical modeling shows a small ($-1.3\%$) correction to the zero-bias $G_A(11\,\text{mK})$. In the modeling, reducing and increasing the ac excitation by a factor of 2 gives corrections of $-0.28\%$ and $-5.5\%$ to the zero-bias $G_A(11\,\text{mK})$, respectively. This is in good agreement with the experimental change of $G_A(11\,\text{mK})$ by approximately $+1\%$ and $-4\%$, compared to 7.2 μV, observed for excitations of 3.6 μV and 14 μV, respectively (all with dc bias $V_{DC} = 0$, *cf.* Section IV). The finite ac excitation correction to the *normalized* $G_A(T)/G_A(11\,\text{mK})$ is calculated to be nearly $T$-independent, changing only by $-9 \times 10^{-4}$ in the experimental temperature range, well below the experimental uncertainty. We therefore conclude that the experimental $G_A(T)/G_A(11\,\text{mK})$ dependence data is essentially in the zero-bias limit.

The experimental $G_A(T)$ dependence is clearly different from the temperature dependence observed in resonant tunneling in quantum dots[34] and antidots,[30,35] and in the Coulomb blockade devices.[36] In resonant tunneling, a single tunneling peak is well described by $G(X,T) = G_0/T \cosh^2(X/T)$, where $X$ is the resonance tuning parameter, even in the FQH regime.[30] The single peak conductance $G_P(T) \propto 1/T$.[35] For the Coulomb blockade, not expected for the nearly open geometry of the interferometer device, a single tunneling peak is given by $G(X,T) = G_0 X/T \sinh(X/T)$. The maximum conductance $G_{MAX}(T) \approx const$ because the single-charge tunneling can proceed via many levels within an energy interval $\propto T$, canceling the $1/T$ Fermi liquid prefactor. In both regimes, the individual $G(X,T)$ tunneling peaks overlap at higher temperatures, resulting in a non-universal $T$-dependence of oscillation amplitude, since it depends on the level spacing $\Delta E$ or the charging energy $U_C$. At arbitrary temperatures, the two $G_A(T)$ dependencies can be evaluated numerically, with an activation temperature $T_0 \propto \Delta E$ or $\propto U_C$ as a fitting parameter.

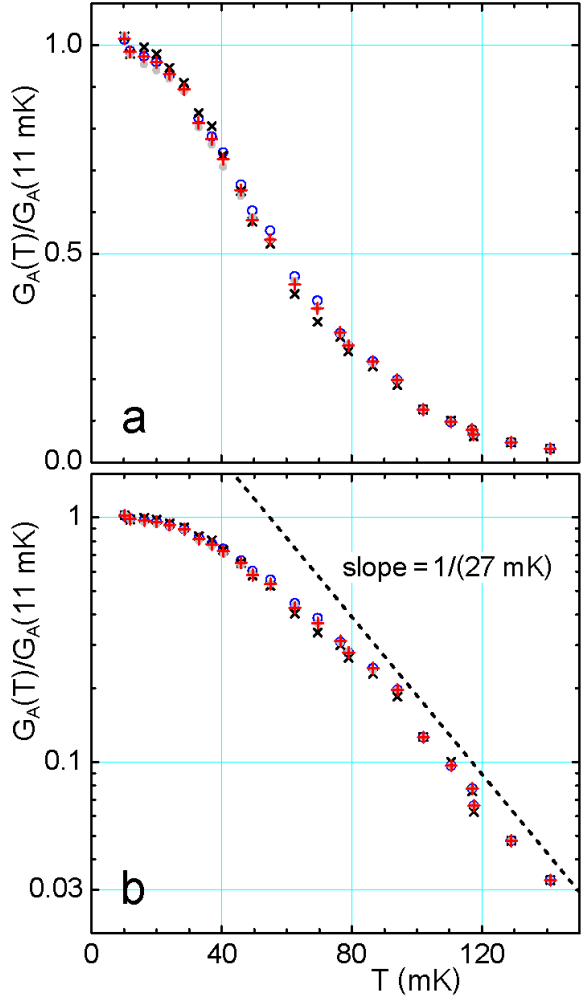

FIG. 3. Temperature dependence of the conductance amplitude of the superperiodic oscillations. Various symbols give data points with one of the oscillations included or excluded from the average, as described in the text; the spread illustrates the uncertainty in the data. The high-$T$ asymptotic behavior $G_A(T) \propto \exp(-T/T_0)$, expected for thermal broadening in a quantum-coherent electron interferometer, is apparent in the semi-log plot (b). The dashed straight line shows the $T_0 = 27$ mK slope.

Figure 4 gives the best fit of the experimental $G_A(T)$ to the resonant tunneling theory with two free parameters: the $T \to 0$ peak conductance $G_0$, proportional to the tunnel coupling $|\Gamma|^2$, and $T_0$. The fit includes electron heating effects by using a finite phenomenological electron heating temperature $T_H$, incorporated as $T_{eff} = (T^5 + T_H^5)^{1/5}$. Specifically, $T_H = 18$ mK is used, obtained for a small quantum antidot on the $f = 1/3$ plateau,[30] where electron heating effects are expected to be more important than in the larger interferometer device. The fit is very sensitive to the value of $T_H$ because of the diverging $1/T$ prefactor; lowering $T_H$ makes the fit worse. We stress that no physically realistic value of $T_H$ produces a good fit to the interferometer data. Fig. 4 also shows the best two-parameter fit of the Coulomb blockade theory. A large deviation of the fit at low $T \ll U_C$ is apparent, with the experimental $G_A(T)$ still rising, while the theoretical dependence saturates at ~32 mK. This deviation is not curable by including electron heating effects. Besides, the Coulomb blockade $T$-dependence results from the conductance minima rising, while the maxima remain nearly constant with increasing $T$, a behavior clearly different from that seen in this sample, Fig. 2(b). Both RT and CB fits give the energy scale ~500 mK, surprisingly large for a ~2 μm lithographic diameter device. Thus we conclude that neither the single-particle resonant tunneling nor the classical Coulomb blockade occurs in the LQP interferometer device.

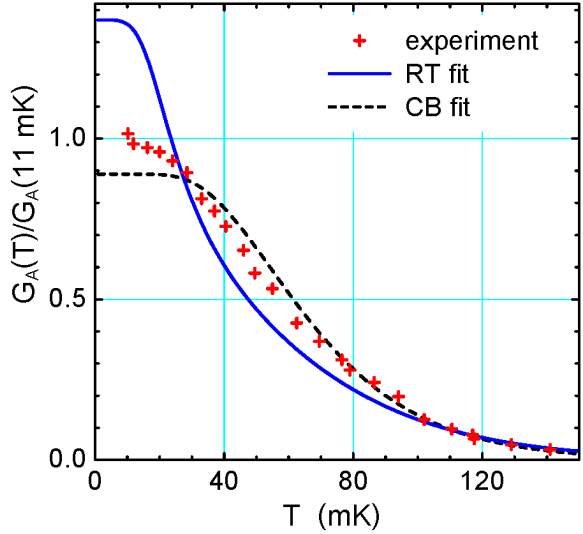

FIG. 4. Two-parameter best fits to the oscillatory conductance behavior expected for single-particle resonant tunneling (RT) and Coulomb blockade (CB) devices. The fits use a phenomenological $T_H = 18$ mK, the electron heating temperature, Ref. 30. The RT fit is very sensitive to the value of $T_H$ because of the $1/T$ Fermi liquid prefactor. The CB oscillation amplitude saturates at ~32 mK, while the experimental amplitude does not.

To elucidate the constraint the LQP interferometer data impose on a possible theoretical model, we fit the data to a phenomenological expression

$$G_A(T) = G_0 (T/T_0)^\alpha [\sinh(T/T_0)]^{-1}, \qquad (1)$$

where the power $\alpha$ is an additional parameter. An example of such a fit with $\alpha = 1$ and $T_H = 15$ mK is shown in Fig. 5. This functional dependence ($\alpha = 1$) is predicted for a chiral Aharonov-Bohm interferometer in the low-bias limit, when the Luttinger exponent $g = 1$.[26] The fit gives $T_0 = 25$ mK, corresponding to the edge excitation velocity $u = 4.0 \times 10^4$ m/s. The value $T_0 = 25$ mK is small, while the oscillations are observable up to ~140 mK. As a comparison, in a quantum antidot in the FQH regime,[30] the corresponding activation $T_0 \approx 120$ mK, while the oscillations were observable up to ~80 mK, a lower $T$ than $T_0$. In the quantum dots[34] and antidots[35] in the integer quantum Hall regime, likewise, conductance oscillations are observable up to ~$0.5 T_0$.

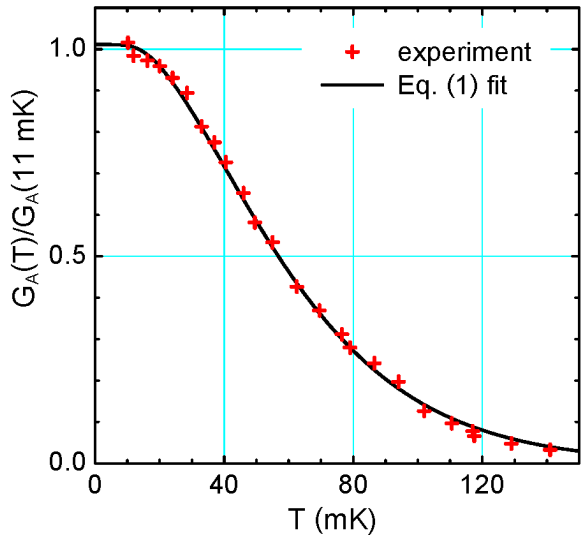

FIG. 5. Fit to Eq. (1) with $\alpha = 1$ and the phenomenological electron heating temperature $T_H = 15$ mK. Eq. (1) with $\alpha = 1$ follows from the chiral Aharonov-Bohm interferometer theory for $g = 1$, Ref. 26.

Satisfactory fits to Eq. (1) are obtainable only with $\alpha \approx 1 \pm 0.3$, limiting $T_H \leq 18$ mK, so that the strictest restriction on the functional $G_A(T)$ dependence in the experimentally-accessible regime is placed by the prefactor to the asymptotic high-$T$ behavior. Neither $\alpha = -1$, as for the Fermi liquid resonant tunneling, nor $\alpha = 5$ predicted for the $g = 1/3$ chiral Aharonov-Bohm effect in a strong-coupling limit in a quantum antidot[27] agrees with the experimental LQP interferometer $G_A(T)$ dependence.

## IV. CHIRAL LUTTINGER LIQUID INTERFEROMETER MODEL

Before we proceed to a quantitative analysis of the temperature dependence, we consider the Hall voltage $V_H$ dependence of the conductance. Experimentally, the applied sample current $I_X$ is controlled, resulting in $V_H \cong 3hI_X/e^2$ on the $f = 1/3$ plateau (for weak tunneling). The Hall (tunneling) current $I_H$ gives rise to the oscillatory $\delta R_{XX}$ directly measured in the experiment. The differential conductance $G(V_H) \approx dI_H/dV_H$ is obtained in the lock-in technique by adding a dc bias to the ac modulation $\delta I_X$, so that $I_X = I_{DC} + \delta I_X$. The experimental differential conductances $G_{MAX}$ and $G_{MIN}$, determined on the maximum and the minimum of a conductance oscillation at the lowest bath $T$, are shown in Fig. 6. The average conductance $G_{AVE} = \tfrac{1}{2}(G_{MAX} + G_{MIN})$ is shown in Fig. 7. The conductance oscillation amplitude $G_A = \tfrac{1}{2}(G_{MAX} - G_{MIN})$ crosses zero, and is plotted in Fig. 8.

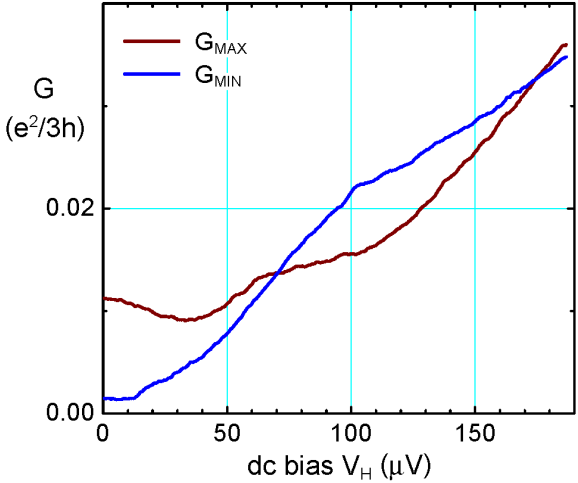

FIG. 6. Experimental differential conductance at the maximum ($G_{MAX}$) and the minimum ($G_{MIN}$) of a $\cos\gamma$ superperiodic oscillation at the lowest bath temperature.

In order to analyze these data, following Ref. 26, we write the total interferometer tunneling current as

$$I_H = I_0 F(V,T)[1 + \cos\gamma\, H(V,T)]. \tag{2}$$

A geometry-dependent $I_0$ is proportional to the $e/3$ LQP tunnel coupling $|\Gamma|^2$, the $\cos\gamma$ term describes the Aharonov-Bohm oscillations (the Berry phase $\gamma = 2\pi\Phi/\Delta_\Phi$), and $F(V,T)$ and $H(V,T)$ are dimensionless functions giving the overall (non-oscillatory) and the oscillatory $I-V$ characteristics of the device, respectively. Differentiation with respect to $V = V_H = R_{XY}I_X$ and setting $\cos\gamma = \pm 1$ for the maximum and minimum conductance give the non-oscillatory conductance ($G_{AVE}$) and the amplitude of the $\cos\gamma$ oscillatory conductance ($G_A$):

$$G_{AVE} = I_0(\partial F/\partial V), \tag{3a}$$

$$G_A = I_0[(\partial F/\partial V)H + F(\partial H/\partial V)]. \tag{3b}$$

Thus, $G_{AVE}$ depends only on $F(V,T)$, while $G_A$ depends on both functions.

Chamon et al. have calculated the Aharonov-Bohm oscillatory conductance expected for a two point-contact interferometer formed by χLL edge channels.[26] Their geometry is similar to ours, the most apparent difference is that they consider only one quantum Hall filling in the interferometer. They obtain analytical functions $F_g(V,T)$ and $H_g(V,T,u)$ for Eq. (2) for the primary Laughlin states. These functions of the Hall voltage and temperature are characterized by the Luttinger exponent $g$; function $H_g$ also depends on the edge excitation propagation velocity $u$. The functions are explicitly given in Eqs. (19-23) of Ref 26. Physically, the function $F_g(V,T)$ describes internal quasiparticle tunneling between two infinitely-long χLL edges, and is identical to the result for only one tunneling constriction.[23] The function $H_g(V,T,u)$, normalized to unity as $T \to 0$, describes the $\cos\gamma$ interferometric signal, and originates in the closed path of length $C$, the interferometer circumference. Chamon et al. show that $H_g(y_1, y_2)$ is a function of two reduced variables: $y_1 = 2\pi\omega_J/\omega_0$ and $y_2 = \omega_J/2\pi T$. The "Josephson frequency" for the charge $q$ quasiparticles is $\omega_J = qV_H/\hbar$, where $V_H$ is the Hall voltage (the tunneling bias), and the "oscillatory frequency" is defined as $\omega_0 = 4\pi u/C$.

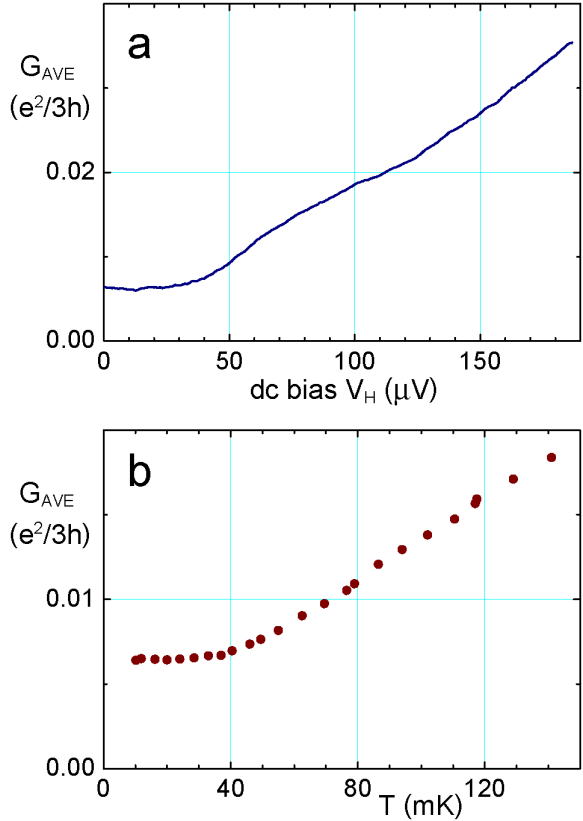

FIG. 7. (a) Experimental average differential conductance $G_{AVE} = \frac{1}{2}(G_{MAX} + G_{MIN})$ of a $\cos\gamma$ oscillation, obtained from the data of Fig. 6 at 11 mK. (b) Experimental average low-bias conductance for six $\cos\gamma$ oscillations as a function of temperature.

The functional dependence $F_g(V,T)$ predicted for $f = 1/3$ FQH constrictions often assumes $g = 1/3$ obtained for the Laughlin wave function.[8,23] $F_{1/3}(V,T)$ is known not to agree well with currently available experiments.[29-31] In the low-$T$, high-$V$ regime, $y_2 > 1$, appropriate for $V_H > 20$ μV in Fig. 7(a), theory predicts $F_g \propto V^{2g-1}$. The predicted low-bias temperature dependence [$T > 20$ mK in Fig. 7(b)] is $F_g \propto V T^{2g-2}$. Thus, assuming $g = 1/3$, the predicted bias dependence for the differential conductance is $G_{AVE} \propto -|V^{-4/3}|$ and the temperature dependence $G_{AVE} \propto T^{-4/3}$, very different from the experimental data in Fig. 7. The experimental $G_{AVE}(V,T)$ is approximately constant for $V_H < 40$ μV and $T < 40$ mK; a relatively moderate *increase* at higher biases and temperatures is in stark contrast to theoretical predictions of a decreasing ($g < 1$) or constant ($g = 1$) $G_{AVE}$. The observed increase in $G_{AVE}$ can probably be attributed to effects not included in theory, such as energy dependence of $|\Gamma|^2$, and a finite sample conduction outside of the interferometer region (becoming noticeable only at higher $T$ and $V$). The predicted $\partial F_g / \partial V$ is nearly temperature-independent for $\omega_J > 2\pi T$, thus perhaps hiding the likely sample heating effects at higher biases.

Theoretically, several physical mechanisms of such discrepancies have been proposed, primarily focusing on effects of long-range Coulomb interaction, edge reconstruction, coupling to a neutral mode, and effects of sample geometry.[24,28,37-39] These effects are modeled in part as renormalization of the Luttinger exponent, $g \to g^* > 1/3$, and can also lead to a modification of the functional dependence of $F_g(V,T)$. Since our primary interest here is in the behavior of the $\cos\gamma$ oscillatory conductance, we therefore simply assume $F_g(V,T) \propto V$, and thus a constant $G_{AVE}$, corresponding to $g = 1$ in the theory of Refs. 23,26.

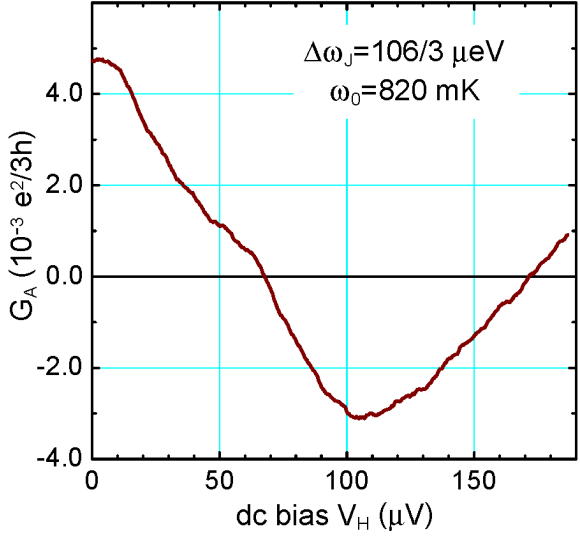

FIG. 8. Experimental differential conductance amplitude $G_A = \frac{1}{2}(G_{MAX} - G_{MIN})$ of a $\cos\gamma$ oscillation, determined from the data of Fig. 6. The separation between the two zero crossings $\Delta\omega_J = e\Delta V_H / 3\hbar$ gives the value of $\omega_0 = 2\Delta\omega_J$.

Our analysis procedure is as follows. We first analyze the experimental oscillatory dc voltage dependence of the differential conductance $G_A(V_{DC})$, Fig. 8, obtaining the value of $\omega_0 = 4\pi u / C$ and thus $u$. We do so because the predicted bias separation between zero-crossings $G_A(V_{DC}) = 0$ is not sensitive to $g$ and $T$ (at low $T$). We then use thus obtained $\omega_0$ to fit the experimental $T$-dependence of the conductance amplitude, using $g$ as the fitting parameter. This analysis procedure is robust because at a low-bias, $\omega_J \to 0$, $\partial H_g / \partial V \to 0$, and the theoretical $G_A = I_0[(\partial F_g / \partial V)H_g + F_g(\partial H_g / \partial V)] \propto |\Gamma|^2 H_g$, using $F_g(V,T) \propto V$, as discussed above. Thus,

the predicted $G_A(T)/G_A(11\,\text{mK}) \propto \beta H_g(T, \omega_0, \omega_J \to 0)$, where a dimensionless scale parameter $\beta \sim 1$ is introduced because the experimental $|\Gamma|^2$ is not known independently. Thus, $\beta H_g$ is not normalized to unity as $T \to 0$. This dependence *approximately* scales with the ratio $\omega_0/g$, see Fig. 9. Shown are the fits of $\beta H_g$ to the data, fixing $g$ and using $\omega_0$ and $\beta$ as the fitting parameters. Hence, the ubiquitous experimental uncertainty does not allow to distinguish between the surprisingly close functional dependencies predicted for various values of $g$ solely from the experimental temperature dependence of $G_A(T)$.

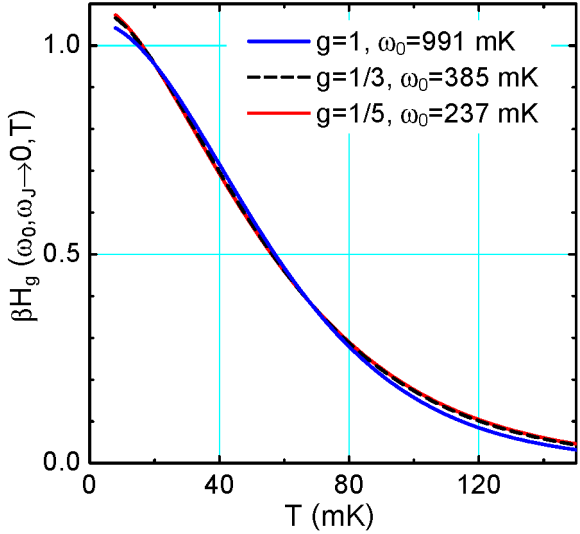

FIG. 9. The temperature dependence of the function describing the interferometric oscillation amplitude for several values of the Luttinger exponent $g$. The fitting parameters are $\omega_0$ and $\beta \approx 1.1$. The function $H_g$ is obtained in Ref. 26; the notation is defined in the text.

Substituting the phenomenological $F_g = G_0 V$ in Eq. (3b), the differential conductance amplitude of the $\cos\gamma$ oscillations is

$$G_A \propto H_g + V(\partial H_g/\partial V). \tag{4}$$

Using the $T \to 0$ theoretical expression $H_g(y_1) \propto J_{g-1/2}(y_1)/y_1^{g-1/2}$, where $J_{g-1/2}(y_1)$ is the Bessel function of the first kind of order $g - \tfrac{1}{2}$, and $y_1 = 2\pi\omega_J/\omega_0$, we obtain

$$G_A \propto y_1^{\tfrac{1}{2}-g} J_{g-\tfrac{1}{2}}(y_1) - y_1^{\tfrac{3}{2}-g} J_{g+\tfrac{1}{2}}(y_1). \tag{5}$$

The separation between the zeros of the Bessel function $J_{g-1/2}(y_1)$ is approximately $\pi$, nearly independent of $g$, while the position of the first zero is moderately sensitive to $g$. Numerical modeling of Eq. (4) shows that the separation between the first two zero crossings is not sensitive to temperature for $T \leq \omega_0/4\pi^2$.

Since $\omega_J = eV_{DC}/3\hbar$ and $\omega_0 = 4\pi u/C$, $y_1 = 2\pi eV_{DC}/3\hbar\omega_0$, the only fitting parameter is $\omega_0$. We therefore fit the separation between the first and the second zero crossings of the experimental $G_A(V_{DC})$, thus obtaining $\omega_0$. This gives $\hbar\omega_0 = 820$ mK, and the chiral edge excitation velocity $u = 3.4 \times 10^4$ m/s. Additionally, we can obtain the effective radial confining electric field at the chemical potential (where the interference path lies): $\mathcal{E} = uB = 4.2 \times 10^5$ V/m.

The approximately circular device has the $f = 1/3$ edge channel circumference $C = 4.0$ μm. These values of the $e/3$ LQP edge velocity and the confining electric field are comparable to the values reported for considerably smaller, but similarly fabricated quantum antidot devices.[30,35] Apparently, the self-consistent screening of the bare depletion potential of the etch trenches by 2D electrons makes the radial electric field weakly dependent on the size of the GaAs/AlGaAs heterojunction device.

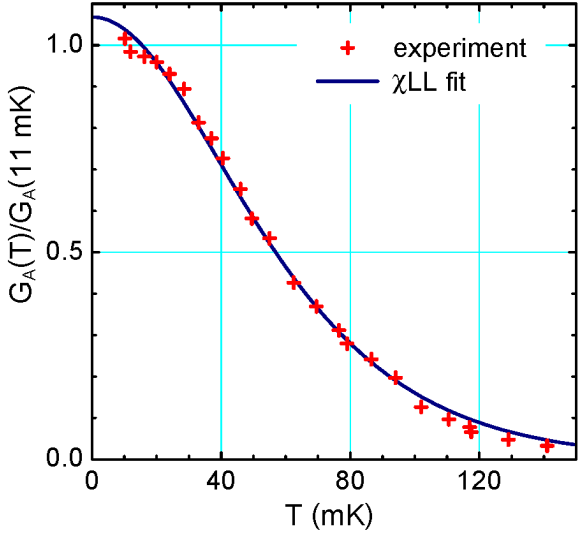

FIG. 10. Fit of Eq. (4) to the experimental data, with the Luttinger exponent $g$ as a fitting parameter. The value $\omega_0 = 820$ mK is fixed by the differential conductance $G_A(V_{DC})$ data, Fig. 8. The best fit is obtained for $g = 0.79$.

At a low bias $\partial H_g / \partial V \to 0$, and Eq. (4) reduces to $G_A \propto H_g(T, \omega_0, \omega_J \to 0)$. We proceed to fit the temperature dependence of the normalized experimental $G_A(T)/G_A(11\,\text{mK})$ to $H_g(u, V \to 0, T)$ of Ref. 26. Since the value of $\omega_0$ is fixed by the differential conductance fit, Fig. 8, the remaining fitting parameter here is $g$. The best fit is shown in Fig. 10. Inclusion of a phenomenological electron heating temperature $T_H = 15$ mK makes the fit slightly better, particularly for the lowest $T < 18$ mK, similar to that in Fig. 5. As discussed above, the theoretical $H_g(\omega_0, \omega_J \to 0, T)$ temperature dependence *approximately* scales with the ratio $\omega_0/g$ for $0.2 \leq g \leq 1$, Fig. 9. Thus, nearly equally good fits of the temperature dependence can be obtained by fixing $g = 1/3$ and using $\omega_0$ as a fitting parameter.[40,41] The low-$T$ absolute value of the conductance amplitude $G_A(11\,\text{mK})$ gives the $e/3$ LQP tunnel coupling $|\Gamma|^2 = \Gamma_1 \Gamma_2^* + \Gamma_1^* \Gamma_2 \approx 0.005$. The small experimental $|\Gamma|^2 \ll 1$ validates our use of the theoretical expression for $H_g(y_1, y_2)$, obtained to the first order in the tunneling amplitudes.[26]

For the present interferometer device, using the experimental values, we can estimate the corresponding $e/3$ single-particle energy level spacing at the chemical potential as $\Delta E \approx \omega_0 / 2\pi^2 \approx 2T_0 \approx 50$ mK. Yet the oscillatory signal persists to $T \sim 6T_0$, which indicates that single-particle energy quantization is not the origin of the oscillatory conductance, and that such quantization is not required for the interferometric signal. As discussed in Section III, the oscillatory conductance in single-particle resonant tunneling and Coulomb blockade devices is experimentally observable only up to $T \sim 0.5T_0$. In these devices, the oscillatory conductance results from the effectively single-particle tunneling dynamics, and thus is suppressed by moderate thermal broadening in the source-drain tunneling electrodes.

## V. TOPOLOGICAL QUBIT PROTECTION

Both theoretical chiral interferometer fits, in Figs. 5 and 10, assume the $e/3$ Laughlin quasiparticles maintain quantum phase coherence while encircling the island, including the tunnel coupling. The only mechanism producing a temperature dependence of the amplitude of interferometric oscillations included in the theory[26] is the thermal broadening of the $f = 1/3$ edge excitations. For example, the theory does not include decoherence resulting from electromagnetic coupling to the metal gate electrodes, nor excitation of edge neutral modes by the tunneling LQPs.[3] Addition of a phenomenological electron heating temperature $T_H$ does not describe quantum decoherence, but an actual heating of the electron system in the sample above the lattice temperature. Thus, the good agreement of the experiment and the theory indicates that the various decoherence mechanisms do not lead to experimentally-observable decoherence.

The chiral interferometer theory of Chamon *et al.*[26] includes no island with filling other than that of the interferometer edge ring itself. At present, it is not known whether inclusion of such island should modify their theory. Such modification seems unlikely, however, on general grounds, since the $e/5$ island quasiparticles do not transport the current and thus should not affect the tunneling dynamics. The interaction of the interfering $e/3$ LQPs with the $e/5$ island quasiparticles is a nonlocal statistical interaction of topological nature.[13,14,33] Such statistical interaction should not affect the dynamics of the interfering $e/3$ LQPs so long as the number of the island quasiparticles remains well-defined, that is, at temperatures $T \ll 2$ K, the 2/5 FQH gap at 12 Tesla.[42]

If the period were not determined by the topological order of the two FQH condensate, one could imagine period changing to $\Delta_\Phi = h/e$ at higher temperatures. Thus, the topological protection of the anyonic statistical interaction is demonstrated experimentally by persistence of the $\Delta_\Phi = 5h/e$ interferometric signal to a relatively high $T = 140$ mK, well above $T_0 \approx 25$ mK. Several designs of qubits and quantum logic gates operating by braiding of anyons in FQH fluids have been proposed.[1-3,16,17] Such topological protection of quantum logic is expected to suppress system decoherence in the appropriately restricted Hilbert space, the primary advantage of fault-tolerant quantum computation with anyons.

## ACKNOWLEDGEMENTS

We thank D. V. Averin, E. Fradkin, E.-A. Kim, S. A. Kivelson and A. V. Shytov for discussions. This work was supported in part by the NSF and by NSA and ARDA through US ARO.

**References**

1. A. Y. Kitaev, Ann. Phys. (N.Y.) **303**, 2 (2003).
2. J. Preskill, in *Introduction to Quantum Computation and Information*, edited by H-K. Lo, S. Papesku, and T. Spiller (World Scientific, Singapore, 1998), pp. 213-269.
3. D. V. Averin and V. J. Goldman, Solid State Commun. **121**, 25 (2002).
4. J. M. Leinaas and J. Myrheim, Nuovo Cimento Soc. Ital. Fis. B **37**, 1 (1977).
5. F. Wilczek, Phys. Rev. Lett. **48**, 1144 (1982); Phys. Rev. Lett. **49**, 957 (1982).
6. M. V. Berry, Proc. R. Soc. London, Ser. A **392**, 45 (1984).
7. D. C. Tsui, H. L. Stormer, and A. C. Gossard, Phys. Rev. Lett. **48**, 1559 (1982).
8. R. B. Laughlin, Phys. Rev. Lett. **50**, 1395 (1983).

9.  For reviews see: *The Quantum Hall Effect*, 2nd Ed., edited by R. E. Prange and S. M. Girvin (Springer, NY, 1990); *Perspectives in Quantum Hall Effects*, edited by S. Das Sarma and A. Pinczuk, (Wiley, NY, 1997); D. Yoshioka, *The Quantum Hall Effect* (Springer, 2002).
10. V. J. Goldman and B. Su, Science **267**, 1010 (1995).
11. F. E. Camino, W. Zhou, and V. J. Goldman, Phys. Rev. B **72**, 075342 (2005).
12. F. E. Camino, W. Zhou, and V. J. Goldman, Phys. Rev. Lett. **95**, 246802 (2005).
13. B. I. Halperin, Phys. Rev. Lett. **52**, 1583 (1984).
14. D. Arovas, J. R. Schrieffer, and F. Wilczek, Phys. Rev. Lett. **53**, 722 (1984).
15. R. B. Laughlin, Rev. Mod. Phys. **71**, 863 (1999).
16. S. Das Sarma, M. Freedman, and C. Nayak, Phys. Rev. Lett. **94**, 166802 (2005).
17. P. Bonderson, A. Kitaev, and K. Shtengel, Phys. Rev. Lett. **96**, 016803 (2006).
18. W. Zhou, F. E. Camino, and V. J. Goldman, Phys. Rev. B **73**, 245322 (2006).
19. R. B. Laughlin, Phys. Rev. B **23**, 5632 (1981); B. I. Halperin, Phys. Rev. B **25**, 2185 (1982); S. A. Trugman, Phys. Rev. B **27**, 7539 (1983); A. H. MacDonald and P. Streda, Phys. Rev. B **29**, 1616 (1984).
20. R. J. Haug, A. H. MacDonald, P. Streda, and K. von Klitzing, Phys. Rev. Lett. **61**, 2797 (1988); S. Washburn, A. B. Fowler, H. Schmid, and D. Kern, Phys. Rev. Lett. **61**, 2801 (1988).
21. P. L. McEuen *et al*., Phys. Rev. Lett. **64**, 2062 (1990); J. K. Wang and V. J. Goldman, Phys. Rev. Lett. **67**, 749 (1991); Phys. Rev. B **45**, 13479 (1992).
22. S. Tomonaga, Prog. Theor. Phys. **5**, 544 (1950); J. M. Luttinger, J. Math. Phys. **4**, 1154 (1963); F. D. M. Haldane, J. Phys. C **12**, 4791 (1979).
23. X. G. Wen, Phys. Rev. B **41**, 12 838 (1990); Int. J. Mod. Phys. B **6**, 1711 (1992).
24. C. de C. Chamon and X. G. Wen, Phys. Rev. Lett. **70**, 2605 (1993).
25. K. Moon *et al*., Phys. Rev. Lett. **71**, 4381 (1993).
26. C. de C. Chamon, D. E. Freed, S. A. Kivelson, S. L. Sondhi, and X. G. Wen, Phys. Rev. B **55**, 2331 (1997).
27. M. R. Geller and D. Loss, Phys. Rev. B **56**, 9692 (1997).
28. A. Lopez and E. Fradkin, Phys. Rev. B **59**, 15 323 (1999).
29. A. M. Chang, Rev. Mod. Phys. **75**, 1449 (2003).
30. I. J. Maasilta and V. J. Goldman, Phys. Rev. B **55**, 4081 (1997); Phys. Rev. B **57**, R4273 (1998).
31. S. Roddaro *et al*., Phys. Rev. Lett. **90**, 046805 (2003).
32. F. E. Camino, W. Zhou, and V. J. Goldman, Phys. Rev. B **72**, 155313 (2005).
33. V. J. Goldman, cond-mat/0605614.
34. L. P. Kouwenhoven, D. G. Austing, and S. Tarucha, Rep. Prog. Phys. **64**, 701 (2001); C. W. J. Beenakker, Phys. Rev. B **44**, 1646 (1991).
35. I. Karakurt, V. J. Goldman, J. Liu, and A. Zaslavsky, Phys. Rev. Lett. **87**, 146801 (2001).
36. D. V. Averin and K. K. Likharev, in *Mesoscopic Phenomena in Solids*, pp. 173-271 (Elsevier, Amsterdam, 1991).
37. L. P. Pryadko, E. Shimshoni, and A. Auerbach, Phys. Rev. B **61**, 10929 (2000); E. V. Tsiper and V. J. Goldman, Phys. Rev. B **64**, 165311 (2001); U. Zulicke, J. J. Palacios, and A. H. MacDonald, Phys. Rev. B **67**, 045303 (2003).
38. A. V. Shytov, L. S. Levitov, and B. I. Halperin, Phys. Rev. Lett. **80**, 141 (1998); X. Wan, K. Yang, and E. H. Rezayi, Phys. Rev. Lett. **88**, 056802 (2002); M. D. Johnson and G. Vignale, Phys. Rev. B **67**, 205332 (2003).
39. B. Rosenow and B. I. Halperin, Phys. Rev. Lett. **88**, 096404 (2002); E. Papa and A. H. MacDonald, Phys. Rev. B **72**, 045324 (2005).
40. A calculation of oscillatory differential conductance at $V_H = 7.5$ μV has been reported by E.-A. Kim, cond-mat/0604359. The theoretical curve fits our data well.
41. F. E. Camino, W. Zhou, and V. J. Goldman, cond-mat/0510764.
42. R. R. Du, H. L. Stormer, D. C. Tsui, L. N. Pfeiffer, and K. W. West, Phys. Rev. Lett. **70**, 2944 (1993).